\documentclass[twocolumn,preprintnumbers,amsmath,amssymb,prb]{revtex4}
\usepackage{graphicx,calc}
\usepackage{bm}
\usepackage{enumerate}
\usepackage[colorlinks=true,dvipdfm]{hyperref}

\begin{document}
\draft
\title{Itinerant Flat-Band Magnetism in Hydrogenated Carbon Nanotubes}

\author{Xiaoping Yang}

\thanks{Corresponding author: xp.yang@fkf.mpg.de}

\affiliation{Max-Planck-Institut f\"ur Festk\"orperforschung, Heisenbergstrasse 1, D-70569 Stuttgart, Germany}

\author{Gang Wu}

\thanks{Corresponding author: wugaxp@gmail.com}
\affiliation{Institute of High Performance Computing, 1 Fusionopolis Way,
16-16 Connexis, Singapore 138632, Singapore}

\date{\today}

\begin{abstract}
We investigate the electronic and magnetic properties of hydrogenated carbon nanotubes using $ab$ $initio$ spin-polarized calculations within both the local density approximation (LDA) and the generalized gradient approximation (GGA). We find that the \emph{combination} of charge transfer and carbon network distortion makes the spin-polarized flat-band appear in the tube's energy gap. Various spin-dependent ground state properties are predicted with the changes of the radii, the chiralities of the tubes and the concentration of hydrogen (H). It is found that strain or external electric field can effectively modulate the flat-band spin-splitting, and even induce an insulator-metal transition.\\


\end{abstract}


\maketitle

\section{Introduction }

Much attention has long been devoted to finding macroscopic
magnetic ordering phenomena in organic materials. Experimentally,
ferromagnetism has been discovered in pure carbon systems, such as
carbon foam, graphite, oxidized C$_{60}$ and polymerized
rhombohedral C$_{60}$ \cite{r1,r2,r3,r4,r5}, which has stimulated
renewed interests in their fundamental importance and potential
applications in high-technology, e.g., the spintronics. Much
theoretical work has been done to study magnetism in nanographites \cite{r6,r7,r8,r9,r10}, C$_{60}$ polymers \cite{r11},
and all-carbon nanostructures \cite{r12,r13}. However, the
microscopic origin of ferromagnetism remains controversial.

The recent proton irradiation experiments in graphite\cite{r2,r3} have shown the importance of hydrogen in inducing
the magnetization instead of magnetic impurities. Disorder
induced by He$^{+}$ ions irradiation does not produce such a large
magnetic moment as obtained with protons \cite{r14}. Theoretically,
the possible magnetism arising from the adsorption of hydrogen atom
on graphite has been studied \cite{r9,r10}. It can be easily
speculated that hydrogen can trigger the $sp^2$-$sp^3$
transformation, promoting the magnetic ordering in other carbon
structures, especially carbon nanotubes with a surface of positive
curvature.

Herein, we focus on the electronic and magnetic properties of
single-walled carbon nanotubes (SWNTs) with hydrogen atoms
adsorbed on their surfaces. Our results show that hydrogenated
carbon nanotubes are on the verge of magnetism instability, and the
\emph{combination} of charge transfer and carbon network distortion
drives flat-band ferromagnetism. To our knowledge, this is
the first comprehensive $ab$ $initio$ study on the physical origin of 
flat-band ferromagnetism in the real carbon nanotube materials.
Moreover, the applied strain and external electric field are found
to have a strong influence on the flat-band spin-splitting,
resulting in the variation of spin-relevant physical properties.

\section{Results and discussion}

\begin{figure}[htbp]
\includegraphics[width=0.5\columnwidth, angle=-90]{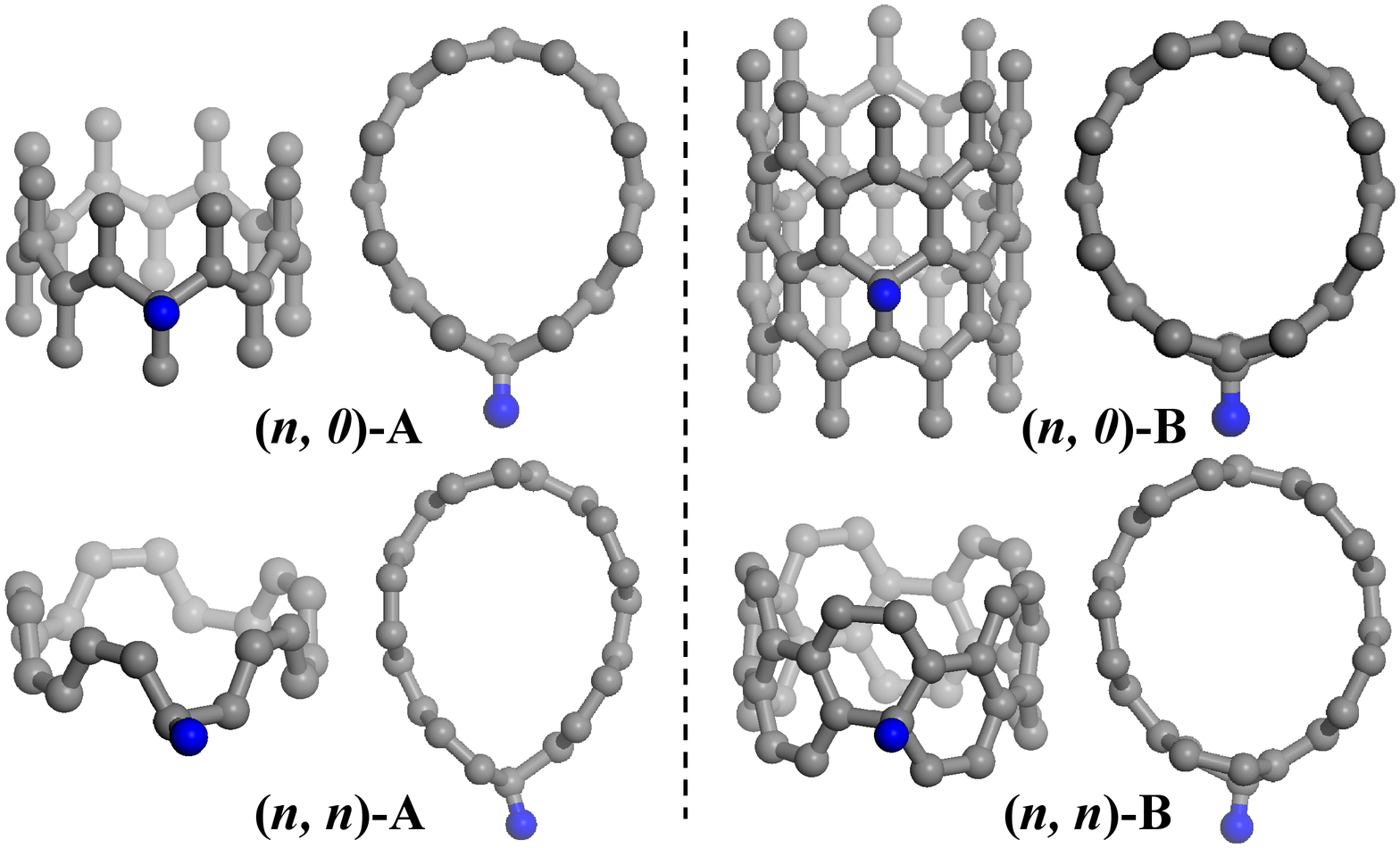}
\label{fig1} \caption{Schematic geometrical structures of 
hydrogenated zigzag (\emph{top panels}) and armchair (\emph{bottom
panels}) tubes in the higher hydrogen concentration A with one hydrogen
per tube period, or the lower hydrogen concentration B with one
hydrogen per every two tube periods. Blue (grey) balls represent the
hydrogen (carbon) atoms.}
\end{figure}

\begin{figure*}[htbp]
\includegraphics[width=1.9\columnwidth]{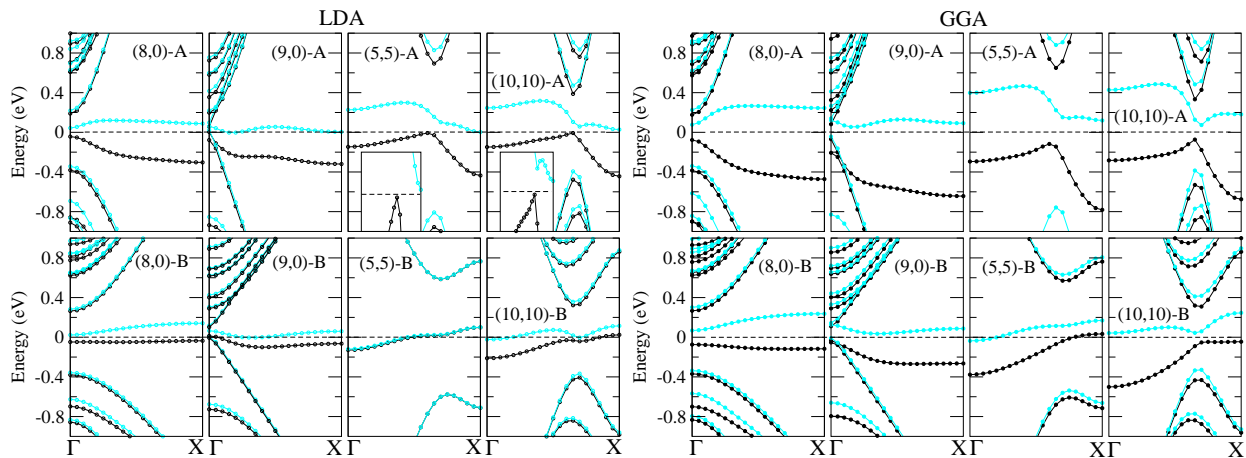}
\label{fig2} \caption{Spin-polarized LDA (\emph{left panels}) and
GGA (\emph{right panels}) ground state band structures of
hydrogenated zigzag (8,0) and (9,0), and armchair (5,5) and (10,10)
tubes in the two different systems A (\emph{top panels}) and B
(\emph{bottom panels}). Insets show the band structures around
$\epsilon _{F}$ in an enlarged energy scale. The Fermi level is set
at zero. Spin-up and spin-down channels are represented by black and
cyan, respectively.}
\end{figure*}

Instead of the simple $sp^2$ bonds in graphite, the bonds in carbon
nanotubes are of $sp^2$-$sp^3$ character due to the tube's curvature
effect, making the hybridization of $\sigma $, $\sigma ^\ast $, $\pi
$ and $\pi ^\ast $ orbitals quite larger, especially for 
small-diameter tubes. Thus the magnetic property of hydrogenated
SWNTs are more complicated than that of hydrogenated graphite,
with a large dependence on the radii, the chiralities, and hydrogen 
concentration. Here, two types of linear hydrogen
concentration A and B are shown in Fig. 1. There is one hydrogen
atom per tube period in the higher concentration A, and one
hydrogen atom per every two tube periods in the lower concentration B. The
higher H concentration leads to a larger structure deformation, as
can be seen in Fig. 1. Stable C-H bond length is 1.1 {\AA}, typical
of covalent bonding (cf. 1.09 {\AA} in methane). Recently, the 
cooperative alignment of the absorbed atoms has been observed in graphene
experimentally \cite{r15}. In the case of carbon nanotubes,
the absorption of hydrogen on tube wall is easier than that on graphene due to the curvature effect \cite{r16}. Moreover, the adatoms' cooperative
alignment can be enhanced by high curvature regions of nanotube
\cite{r16}, that can result from pressure \cite{r17}, compression
transverse to its axis \cite{r16,r18}, or tube-substrate interaction
\cite{r19}.

Now, let us study electronic structures of hydrogenated SWNTs,
taking the zigzag (8,0), (9,0), armchair (5,5) and (10,10) tubes as
examples. Fig. 2 presents their ground state band structures
obtained by using spin-polarized LDA (\emph{left panels}) and GGA
(PBE exchange correlation functional) (\emph{right panels}). In
spin-unpolarized LDA and GGA paramagnetic (PM) band structures (not
shown here completely), a common feature is that hydrogen atom
induces a half-filled flat-band in the tube's energy gap around
$\epsilon _{F}$ due to the odd electrons in compounds, as seen in 
the LDA-PM ground state band structure of (5,5)-B in Fig. 2.
Apparently, flat-band causes an extremely high density of states
around $\epsilon _{F}$, and if the Coulomb interaction between
itinerant electrons in the band is introduced, magnetic instability would occur. It has been shown that flat-band leads to ferromagnetism for certain models \cite{r20}. In hydrogenated SWNTs, the spin-spin interaction plays a
similar role as the Coulomb one, and lifts the spin degeneracy of the flat
band. As a result, the flat-band's spin-splitting magnitude and the
energy position relative to $\epsilon _{F}$ determine the 
ground state properties of system. As we can see in the right panels of
Fig. 2, the spin-splitting magnitude has increased after introducing
the generalized gradient correction in the exchange correlation
functional, compared to the LDA without correction (\emph{left
panels}), which makes ferromagnetic (FM) state become more favorable in
energy. The adsorption of hydrogen atom is found to hardly affect
the gaps of zigzag (8,0) and (9,0) tubes \cite{r21}, while a large tube's
energy gap is opened in the metallic armchair (5,5) and (10,10) tubes. In the LDA
results, (8,0)-A,B, (5,5)-A and (10,10)-A exhibit FM semiconducting characteristic, whereas
(9,0)-A,B and (10,10)-B are FM metals. (5,5)-B has a
PM metallic ground state under LDA. However, the enhanced
spin-splitting in GGA makes (9,0)-A,B and (10,10)-B present a FM
semiconducting behaviour, not metallic one under LDA. Furthermore, a FM metallic state is produced for (5,5)-B, not the PM metallic one obtained by LDA.\\

\begin{figure}[htbp]
\includegraphics[width=0.8\columnwidth]{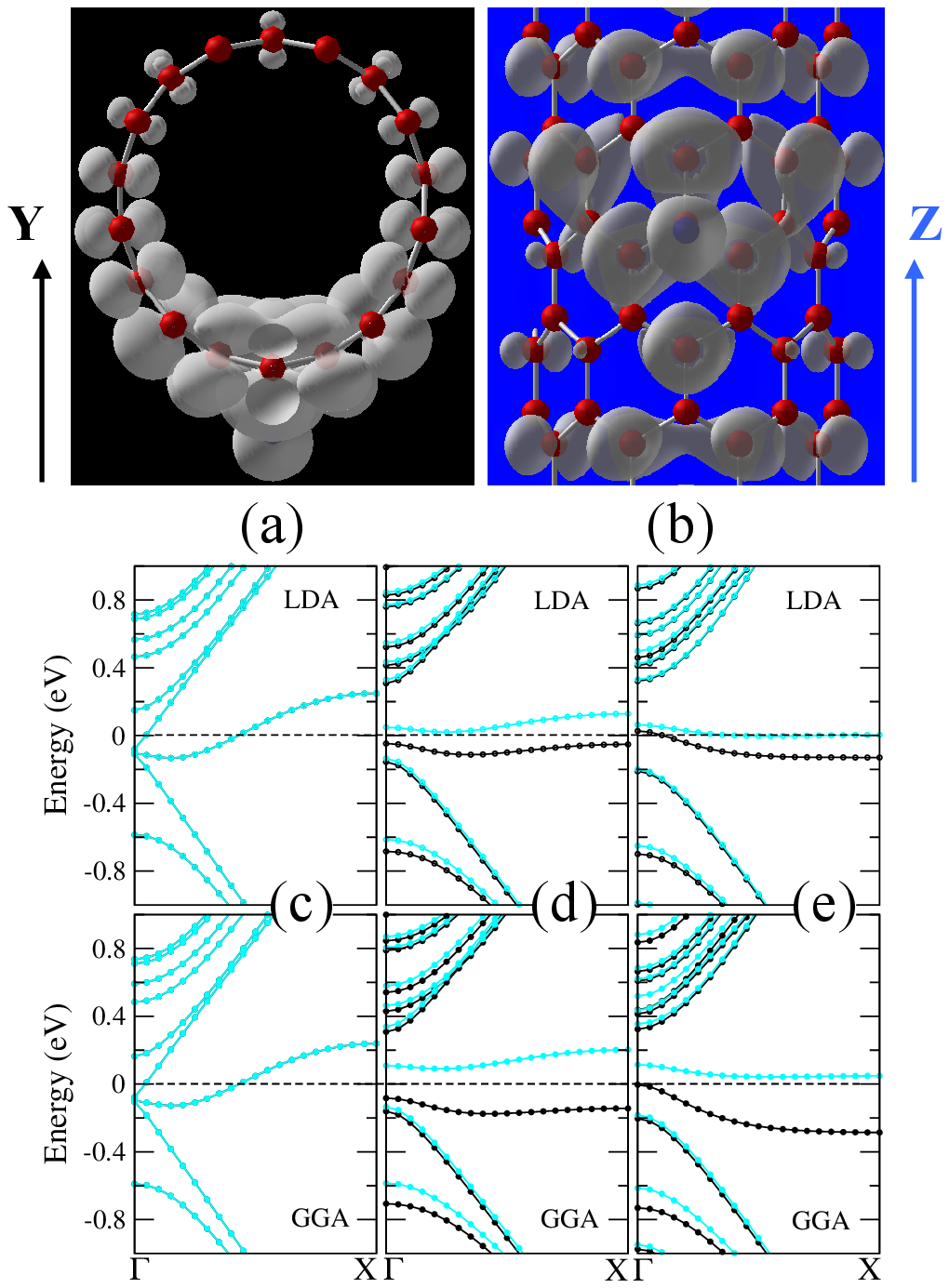}
\label{fig3} \caption{The 0.003 {\AA}$^{-3}$ magnetization density
isosurfaces for the full-occupied spin-up GGA flat-band in (9,0)-B:
(a) top view, and (b) front view. Spin-polarized LDA and GGA ground
state band structures for (9,0)-B: (c) with only the location of
hydrogen atom relaxed, (d) under the $4\%$ axial stretch strain, (e)
under the $4\%$ axial compression strain.}
\end{figure}

In order to investigate the physical origin of flat-band
ferromagnetism, we plot the spin-up density of the full-occupied GGA flat-band of (9,0)-B (black band, see Fig. 2) in Fig. 3.
A $sp^3$-like hybridization is induced on the carbon atom attached
to the hydrogen atom, leading to whole carbon network distortion,
accompanied by charge transfer from hydrogen to the bonded carbon
atom (0.352 and 0.334 electron for LDA and GGA (PBE), respectively).
Magnetism is strong itinerant in both circumferential and tube
axial directions, arising from the hybridization of H-$s$ orbital
with tube-$\pi$ orbitals. Magnetic moment per unit cell is 0.63 and
0.96 $\mu _B $ in LDA and GGA, respectively.

Obviously, this $s$-$\pi$ hybridization is relevant to
structure distortion and affects the flat-band spin-splitting
magnitude. To get such an insight, we carried out both LDA and GGA calculations on the
undistorted (9,0)-B structure with only the location of hydrogen
atom relaxed. Charge transfer still exists, and the band structures
plotted in Fig. 3(c) clearly show a disappearance of the flat-band
spin-splitting and also magnetism. Weak magnetic moment 0.2 $\mu_B $
 per unit cell appears even if on-site Coulomb repulsion U = 3.0 eV \cite{r22} is considered in the LDA+U calculation. These results
indicate that only charge transfer is not sufficient to induce a
large magnetic moment, therefore structure distortion is indispensable. As
a result, we can anticipate that the applied strain could tune the
physical properties of hydrogenated SWNTs by affecting the
spin-splitting of flat-band. We applied a $4\%$ stretch or compression
strain along the tube axis, then the atomic positions were
optimized again. Corresponding ground state electronic structures
are given in Figs. 3(d)-3(e). As expected, strain do has an important
effect on the magnetic ground state properties. The $4\%$ axial stretch strain
causes the length of C-C bonds, beneath the H atom, further stretched
and deviate from the regular value of 1.42 \AA, which enlarges the energy gap, as shown in Fig. 3(d). Even a metal-insulator transition
is found in LDA band structure, accompanied by an increase in the magnetic
moment to 0.96 $\mu _B$ (normally 0.63 $\mu _B$). Under $4\%$
axial compression strain, the compound presents the metallic or semiconducting
characteristic similar to those without strain in both LDA and GGA, but the magnetic
moment in LDA is reduced to 0.53 $\mu _B $. Under both axial strains, the
transferred charge remains to be 0.35 $\sim $ 0.36
electron in LDA and 0.33 $\sim $ 0.34 electron in GGA, proving that
carbon network distortion plays a very important role in
the flat-band spin-splitting. Of course, flat-band and magnetism would
also disappear if we remove the hydrogen atom from the distorted
SWNTs, indicating the necessity of hydrogen in inducing the
ferromagnetic ordering, which is similar to the role of so called
``carbon radicals" in magnetic all-carbon structures \cite{r7}.

Fig. 4 summaries structure information and ground state properties
of hydrogenated zigzag ($n$,0) ($n$=6--12) and
armchair ($n$,$n$) ($n$=5--11) tubes in both A and B cases (see
Fig. 1). Generally speaking, magnetic moment per unit cell has
increased gradually with the change of tube index $n$ (\emph{top
panels}), and approaches a saturation value in both LDA and GGA
results. In ($n$,0)-A case, magnetic moment curve presents
oscillation behavior (especially in LDA) that might be relevant
to periodic change of the band gap in zigzag-type tubes. The higher
concentration A-type compounds have a larger magnetic moment than
that of the lower concentration B-type compounds, especially in 
smaller diameter tubes. In the bottom panels of Fig. 4, the curves
of energy difference ($E_{PM}-E_{FM}$) between FM and PM states per unit
cell vs tube index $n$ present a similar trend as found in the
curves of magnetic moment vs $n$. Introducing of the generalized
gradient correction enhances the spin-splitting, as discussed for
Fig. 2, which leads that the FM state becomes more favorable
in GGA than in LDA, evidenced by the energy difference $E_{PM}-E_{FM}$.
These reveal that magnetic properties of hydrogenated
carbon nanotubes are affected by the structure characteristic of the SWNTs host, and the concentration of hydrogen.

\begin{figure}[htbp]
\includegraphics[width=0.9\columnwidth]{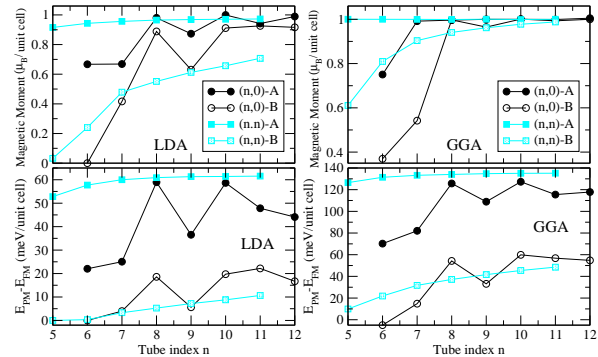}
\label{fig4} \caption{From the top down, the evolution of magnetic
moment per unit cell and energy difference
($E_{PM}-E_{FM}$) per unit cell vs the tube index $n$. Left
panels are from spin-polarized LDA, and right panels are from
spin-polarized GGA (PBE).}
\end{figure}

Finally, we also investigate the effect of transverse electric
fields in the Y direction of cross section (see Fig. 3(a)). Absorption of H atom makes SWNT become a polar organic compound. Here, the polar C-H bond is just along the Y
direction. It can be anticipated that charge redistribution driven
by an external electric field could occur, which would induce a
subtle change of electronic structure. Both LDA and GGA simulations
on (10,10)-A reveal the spin-dependent effect of applied external
electric fields with a magnitude of 0.1 $\sim$ 0.4 V/{\AA}, illustrated in Fig. 5. Compared with Fig. 2, the applied
field gradually decreases the gap between spin-up and -down flat band at 0.1 and 0.2 V/{\AA}, finally closes it and directly
drives an insulator-metal transition at 0.3 V/{\AA} in LDA and 0.4
V/{\AA} in GGA. This effect is different from the case of graphene nanoribbons \cite{r6}, where the band gap closure takes
place only for one spin channel. The fantastic ``beating" behaviour
of spin-polarized flat-band under the external electric field can be
used for designing quantum switches.

\begin{figure}[htbp]
\includegraphics[width=0.9\columnwidth]{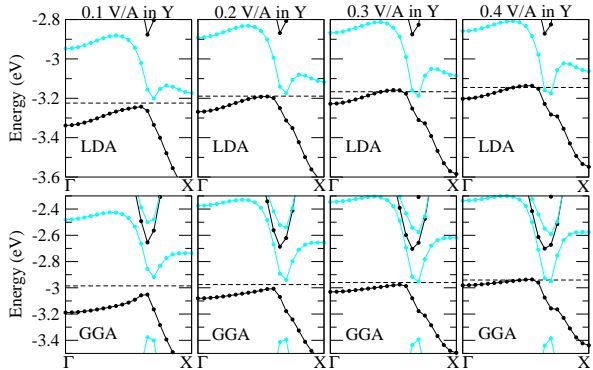}
\label{fig5} \caption{Spin-polarized LDA (\emph{top panels}) and GGA
(\emph{bottom panels}) ground state band structures of hydrogenated
armchair (10,10)-A with the applied electric fields in the modulus
of 0.1$\sim$0.4 V/{\AA} along the Y direction of cross section
(see Fig. 3(a)). The Fermi Levels are marked in black dash lines.}
\end{figure}

\section{Conclusions}

We have demonstrated how the \emph{combined effect} of charger
transfer and carbon network distortion makes spin-polarized
flat-band appear in the tube's energy gap, based on spin-polarized LDA and GGA (PBE) calculations, which is the origin of a variety
of spin-relevant physical properties. Furthermore, hydrogenated SWNTs' ground
state properties are found to depend largely on the radii, the chiralities of the
SWNTs and the concentration of hydrogen. Applied
strain and transverse electric field can effectively tune the
flat-band spin-splitting strength, and even induce insulator-metal
transition. Our results indicate that hydrogenated
SWNTs are the promising candidates for the carbon-based
nanometer-ferromagnetism, spintronic devices, and even quantum switches.
Also, the intrinsic ferromagnetism mechanism revealed in our paper
can be applied to hydrogenated graphene and other carbon-based material
with extended surface.\\

\section{Theoretical methods and Models}

We carried out the numerical calculations using the Vienna $ab$
 $initio$ Simulation Package (VASP) \cite{r23} within the frameworks of spin-polarized local density approximation (LDA) and generalized gradient approximation (GGA) (PBE exchange correlation functional) \cite{r24}. The ion-electron
interaction was modeled by the projector augmented wave (PAW) method
\cite{r25} with a uniform energy cutoff of 400 eV. We used periodic
boundary conditions and a supercell large enough to prohibit the electronic and the dipole-dipole interactions between neighboring tubes. The spacing between $k$ points was 0.03 {\AA}$^{-1}$: 1$\times$1$\times$8 k-point sampling in ($n$,0)-A, 1$\times$1$\times$4 in ($n$,0)-B, 1$\times$1$\times$14 in ($n$,$n$)-A, and 1$\times$1$\times$7 in ($n$,$n$)-B. The geometrical structures of hydrogenated SWNTs were optimized by employing
the conjugate gradient technique, and in the final geometry no
force on the atoms exceeded 0.01 eV/{\AA}.


\end{document}